\documentstyle[preprint,aps]{revtex}
\preprint{IMSc/95-96/001}
\begin{document}
\draft
\title{Solar and atmospheric neutrino oscillations with three
flavours}
\author
{ Mohan Narayan$^a$, M.V.N.Murthy$^a$, G. Rajasekaran$^a$ and S. Uma
Sankar$^{a,b}$}
\address
{$^a$Institute of Mathematical Sciences, Madras 600 113, India.\\
 $^b$Department of Physics, Indian Institute of Technology, Bombay}
\date{\today}
\maketitle

\begin{abstract}
We analyze the solar and the atmospheric  neutrino problems in the context
of  three flavour
neutrino oscillations. We assume a mass hierarchy in the vacuum mass
eigenvalues $\mu_3^2 \gg \mu_2^2 \geq \mu_1^2$, but make no
approximation regarding the magnitudes of the mixing angles. We find
that there are small but continuous bands in the  parameter space where
the constraints imposed by the current measurements of $ \ {}^{71} Ga$,
${}^{37} Cl$ and Kamiokande experiments are satisfied at $1 \sigma$ level.
The allowed parameter space increases dramatically if the error bars are
enlarged to $1.6 \sigma$.
The electron neutrino survival probability has different energy dependence
in different regions of the parameter space. Measurement of the recoil
electron energy spectrum in detectors that use $\nu - e$ scattering may
distinguish between some of the
allowed regions of parameter space. Finally we use the results for the
parameter space admitted by the solar neutrinos as an input for the
atmospheric neutrino problem and show that there exists a substantial
region of parameter space in which both problems can be solved.
\end{abstract}
\vspace{0.5cm}
\pacs{PACS numbers:  14.60.Gh, 96.60.Kx, 95.30.Cq, 96.40.Tv}
\narrowtext

\section{Introduction}

The solar neutrino problem has been an interesting and intriguing
phenomenon in neutrino physics for a long time. The different solar neutrino
experiments observe differing fractions of the neutrino flux predicted by the
Standard Solar Model (SSM) \cite{kvls94,pgl94}.
The oldest of the solar neutrino experiments is the ${}^{37} Cl$ experiment
at Homestake. Its energy threshold is $0.814$~MeV and it can detect the
neutrinos from ${}^7 Be$ $(E_{\nu} = 0.862$~MeV) and ${}^8 B$ $(E_{\nu}
\leq 14.02 \ {\rm MeV})$ reactions.
In the standard solar model (SSM) of Bahcall-Pinnsonneault \cite{bahpin92},
the capture rate in the $^{37}Cl$ experiment is predicted to
be $8.0 \pm 1.0$~SNU. However, the measured rate is only
\cite{homest94}
\begin{equation}
R_{Cl} = 2.55 \pm 0.25 \ {\rm SNU}. \label{eq:cl37}
\end{equation}
The water Cerenkov detector at Kamioka, with a threshold
of $7.5$~MeV, can detect only the neutrinos from the upper end of
${}^8 B$ spectrum and the Kamioka result \cite{kamiok94} is,
\begin{equation}
y_{Kam} = \frac{R_{Kam}}{R_{Kam:SSM}} = 0.51 \pm 0.07,
\label{eq:kamio}
\end{equation}
which is the ratio of the observed neutrino flux to that predicted by the
SSM. The gallium experiments SAGE and GALLEX, with energy threshold of
$0.233$~MeV can detect the neutrinos coming from the dominant $p-p$
reaction $(E_{\nu} \leq 0.42 \ {\rm MeV})$ as well as the neutrinos from
${}^7 Be$ and ${}^8 B$ reactions. Their measured rates are
\cite{sage94,gallex94}
\begin{eqnarray}
R_{SAGE} & = & 69 \pm 11 \pm 6  \ {\rm SNU} \nonumber \\
R_{GALLEX}& = & 79 \pm 10 \pm 6  \ {\rm SNU} \nonumber
\end{eqnarray}
and the average is
\begin{equation}
R_{Ga;avg} =
 74 \pm 8  \ {\rm SNU} \label{eq:ga71}
\end{equation}
as opposed to the SSM prediction of $131.5$~SNU.

A rough model independent analysis of these results indicates that the low
energy  neutrinos from the $p-p$ reaction suffer very little suppression
whereas the higher energy neutrinos are suppressed to a large extent
\cite{kvls94,nhpl94b}. Recently it was pointed out that if neutrinos
have no properties beyond those in the standard electro-weak model (i.e.
if  they are massless), the measurement of Kamiokande, together with that
of ${}^{37} Cl$ experiment, implies that the ${}^7 Be$ neutrinos
must be suppressed by more than $90 \%$ \cite{rosen94,parke95}.

In addition there exists an anomaly in the ratio of observed muon
neutrinos to electron neutrinos in the earth's atmosphere.  These
neutrinos are produced from the decay of $\pi^{\pm}$ and $K^{\pm}$
which are in turn produced by cosmic rays interacting with the
atmosphere.  The
ratio is roughly two as suggested by the Monte-Carlo calculations whereas
both Kamioka\cite{kamat94} and IMB\cite{imbat} report that the ratio is
only about half of that
predicted by the Monte-Carlo calculations. The results for this ratio are
also available from three other groups using the tracking detectors, namely
the NUSEX\cite{nusex}, Frejus\cite{frejus} and SOUDAN-II\cite{soudan}
collaborations.  The data from the NUSEX collaboration seems to be in
agreement with no-anomaly situation. Similar conclusion is obtained from
Frejus data if all the contained events are considered.  However, if only
fully contained results are taken into consideration, there is a
suppression. The SOUDAN-II results are consistent with the
results obtained with water Cerenkov detectors.  It should
be noted that the statistics in the tracking experiments is not as
high as the water Cerenkov experiments.
Evidently any solution of the solar neutrino puzzle must incorporate
simultaneously a solution of the atmospheric neutrino
problem\cite{alp93}.

A satisfactory solution to the solar neutrino problem should
be  able to explain not only the total deficit that is observed but the
differential suppression observed at low and high energies.
Solutions based on astrophysics or nuclear physics ascribe the deficit to
smaller solar core temparature or smaller cross sections for the
nuclear reactions talking place in the sun.
Recent model independent analyses suggest that
these solutions cannot describe the results of ${}^{37} Cl$ and
Kamiokande simultaneously \cite{pgl94,nhpl94b}. Particle physics based
solutions attempt to account for the deficit by assuming that the
neutrinos have interactions beyond those of the
standard electro-weak model. If the neutrinos possess small mass, an
electron
neutrino can oscillate into a neutrino of another flavour \cite{smbbp78}.
The amplitude of oscillation is a function of the mass squared
differences, the mixing angles between neutrino flavours
and the neutrino energy.
%For mass square differences of the order of ${\rm eV}^2$,
%the mixing angles are constrained to be small \cite{numixexp}.
If one of the mass square differences is of the order of the
effective mass squared arising from $\nu_e-e$ interaction,
the matter effects can
enhance the mixing to its maximal value and the
amplitude for $\nu_e$ oscillating into another flavour will be
very large \cite{w79ms85}. This is the so called MSW effect.

Matter-enhanced oscillations have been studied thoroughly in the
scenario where only two flavours, $\nu_e$ and $\nu_{\mu}$,
mix with each other \cite{bethe86,hbjb90,nhpl94a}.
The vacuum oscillation here is controlled by the two parameters,
the mass square difference $\delta_{21} = m_2^2 - m_1^2$
and the mixing angle $\omega$. Matter effect is taken into account
by adding to the mass squared of $\nu_e$, the term
\begin{equation}
A(r) = \sqrt{2}~ G_F ~n_e (r) \times 2 E, \label{eq:defA}
\end{equation}
which is proportional to the electron number density
in the Sun $n_e (r)$, where $r$ is the radial distance from the centre of
the Sun. The maximum value of $A$ occurs at the core and
is roughly $10^{-5} E \ {\rm eV}^2$, where $E$ is the neutrino energy in
MeV. The mixing angle $\omega_m$ in the presence of matter is given by,
\begin{equation}
\cos 2 \omega_m = \frac{\delta_{21} \cos 2 \omega - A}{
\sqrt{(\delta_{21} \cos 2 \omega - A)^2 + (\delta_{21} \sin 2 \omega)^2}}.
\end{equation}
The MSW resonance condition is,
\begin{equation}
A = \delta_{21} \cos 2 \omega. \label{eq:rescon2}
\end{equation}
Note that, if the resonance condition  is to be satisfied, $A_{core} >
\delta_{21}
\cos 2 \omega$, which implies that
$\omega_m > \pi/4$ at core. At resonance it
becomes $\pi/4$ and approaches its vacuum value after passing through the
resonance.

The probability for an electron neutrino produced in the solar core
to be detected as an electron neutrino on earth, averaged over the
time of emission and the time of absorption, is given by
\begin{equation}
\langle P_{ee} \rangle =
\cos^2 \omega \cos^2 \omega_m + \sin^2 \omega \sin^2 \omega_m
 - x_{12} \cos 2 \omega \cos 2 \omega_m, \label{eq:pee2}
\end{equation}
where $\omega_m$  is to be evaluated at the point of production and
$x_{12}$  is the probability of a
non-adiabatic jump between the matter dependent mass eigenstates.
If the variation of the solar density in the resonance region is
slow enough, the adiabatic condition
\begin{equation}
\gamma \equiv \frac{\delta_{21}}{E |\frac{1}{A} \frac{d A}{d r}|_{res}}
\frac{\sin^2 2 \omega}{\cos 2 \omega} \gg 1
\label{eq:adbcon}
\end{equation}
is satisfied and the matter dependent mass eignestates evolve
adiabatically and there are no transitions between them.
If~(\ref{eq:adbcon}) is not satisfied, then there will
be non-adiabatic transitions between the two matter dependent mass
eigenstates
in the resonance region and the probability of this jump has the
general form $exp (-C/E)$. The term $C$ has dimensions of energy and
is some function of $\delta_{21}$, $\omega$ and the derivative of the
solar density. The  expressions for $C$ for various density profiles
are
tabulated in Ref.\cite{kuopanrmp}. For linear density variation in the
resonance region, the jump probability is given by the Landau-Zener
formula
\begin{equation}
x_{12} = \exp \left[ - \frac{\pi}{2} \gamma \right].
\end{equation}

The predictions for the rates of various experiments are
obtained by convoluting the SSM neutrino fluxes with the
expression for survival probability in~(\ref{eq:pee2}).
A fit to the data from ${}^{71} Ga$, ${}^{37} Cl$ and Kamiokande
experiments yields solutions in two regions in the $\delta_{21} - \sin^2
2 \omega$ plane,
one with small vacuum mixing and one with large vacuum mixing:
\begin{eqnarray}
\delta_{21} \simeq 6.1 \times 10^{-6} {\rm eV}^2 & ~{\rm and}~ &
\sin^2 2 \omega \sim  0.0065, \nonumber \\
\delta_{21}  \simeq 9.4 \times 10^{-6} {\rm eV}^2 & ~{\rm and}~ &
\sin^2 2 \omega  \sim  0.62.
\end{eqnarray}
In case of the small mixing angle solution, the resonance occurs
for neutrinos with energy greater than $0.6$ MeV. Therefore, the
$p-p$ neutrinos (whose maximum energy is  $0.42$ MeV)
are unaffected whereas the neutrinos with energy greater than
$0.6$ MeV are almost completely converted into $\nu_{\mu}$.
But the measurement of Kamiokande shows that the neutrinos with
energy greater than $7.5$ MeV are suppressed by only a factor of $0.5$.
This can be accommodated through the non-adiabatic jump $x_{12}$
in~(\ref{eq:pee2}). If $C \simeq 10 {\rm MeV}$, or equivalently
$\delta_{21} \sin^2 2 \omega \sim 4 \times 10^{-8} \ {\rm eV}^2$,
then $x_{12}$
is negligible for energies less than $5$ MeV, but becomes appreciable
at higher energies and
$\langle P_{ee}\rangle$ satisfies Kamiokande constraint. The energy
dependence of $\langle P_{ee} \rangle$ in this case is precisely of
the form that is required to satisfy the data from the three solar
neutrino experiments. In the case of the large angle solution, the
non-adiabatic effects are totally negligible and the $\langle P_{ee}\rangle$
is about $0.55$ below $0.5 \ {\rm MeV}$ and slowly falls to about
$0.35$ around $5 \ {\rm MeV}$ after which it it remains
almost independent of the neutrino energy.

In the case of two flavour oscillations, the area of the parameter
space, which can satisfy all the three constraints at $1 \sigma$
level, is very small.
Especially, in the case of the small angle solution, the
requirement that the resonance should occur around $0.6 \ {\rm MeV}$
uniquely fixes the value of $\delta_{21}$.
The requirement that the ${}^7 Be$ neutrinos should be completely
suppressed
and that the high energy ${}^8 B$ neutrinos should have a suppression
of about $0.5$ determines the product
$\delta_{21} \sin^2 2 \omega$ almost exactly.
Therefore, there is very little leeway in the allowed values of
$\delta_{21}$ and $\sin^2 2 \omega$.
An appreciable region of parameter space is allowed only at
$95 \%$ C.L. ( or 2.4~$\sigma$~level). In addition, this simple picture is
inadequate to
simultaneously explain the solar and atmospheric neutrino deficits since
the mass squared differences required are in vastly different regimes.
To explain the atmospheric neutrino anomaly on the basis of
two-flavour vacuum oscillations, one requires a mass squared
difference of the order of $10^{-1}~-~10^{-3}~ eV^2$, with a large
mixing angle.   This must be compared with the best fit to the data in
the case of solar neutrino problem given in eq.(10).
Therefore one has to necessarily consider the scenerio in which all the
three neutrinos participate.  This of course is also a more realistic
situation since the LEP experiments have already pinned down the number
of light neutrino generations to be three.

Three flavour oscillations were considered previously
\cite{kujp87,ajmm88,kuopan}.
However, the uncertainties in the Gallium experiments have come down
significantly in recent times and the parameter region allowed by
the current data will be much smaller.
Recently Joshipura and Krastev\cite{ajpik} have attempted a complete
solution of the
solar and atmospheric neutrino problems in the three generation
frame work.  They present a combined analysis of these two problems in
the framework of the MSW effect and indeed show that there exists a
parameter space in which both sets of data can be reconciled.  Kim and
Lee\cite{kimlee} analyse
these two problems and present a solution based on maximally mixed( in
vacuum) three generations of neutrinos.  This later analysis is however
a rather fine tuned solution since the parameter space allowed is rather
tiny.

In this paper, we analyze the solar neutrino problem by considering the
oscillations between the three neutrino flavours. The analysis is done
with no particular model of neutrino masses and mixings assumed. The
analysis is similar in spirit to that of Joshipura and
Krastev\cite{ajpik}.
We carry
their analysis further and not only map out the full parameter space, but
also discuss the average survival probability and recoil electron spectrum.
In addition we also discuss a non-standard solution where no resonance
occurs but nevertheless there is a parameter space in which all the three
experiments discussed earlier can be reconciled. We also do not make any
assumption about the evolution being adiabatic and take into account
non-adiabatic effects.  These effects may be ignored, however, in parts of
allowed parameter space. In the three generation case the neutrino
oscillations are determined by two mass differences and three mixing
angles neglecting the CP-violating phase. One of the mixing angles is
irrelevant for solar neutrino problem \cite{kujp87,ajmm88} while being
relevant to the atmospheric neutrino problem and
one of the mass differences is
constrained by the atmospheric neutrino deficit. Therefore the solar
neutrino oscillations in the three flavour case are dependent on three
parameters.  Because of the additional parameter, a larger region of the
parameter space is allowed by the solar neutrino data compared to the
two generation scenario. In section 2, we
present the theoretical frame work for our analysis of the solar
neutrino problem  and in section 3, we
present the numerical results for the  solar neutrino problem in
conjunction with the atmospheric neutrino problem. The last section
consists of a brief summary and discussion.

\section{Three neutrino oscillations in matter- a perturbative
analysis}

In this section we discuss the mixing between three flavours of
neutrinos and obtain the probability for a $\nu_e$ produced in the
sun to be detected as a $\nu_e$ on earth. The three flavour eigenstates
are related to the three mass eigenstates in vacuum through a unitary
transformation,
\begin{equation}
\left[ \begin{array}{c} \nu_e \\ \nu_{\mu} \\ \nu_{\tau}
\end{array} \right] = U^v
\left[ \begin{array}{c} \nu_1^v \\ \nu_2^v \\ \nu_3^v
\end{array} \right],
\end{equation}
where the superscript $v$ on r.h.s. stands for vacuum.
The $3 \times 3$ unitary matrix $U^v$ can be parametrized by three Euler
angles $(\omega, \phi, \psi)$ and a phase. The form of the
unitary matrix can therefore be written in general as,
$$ U^{v} = U_{phase}\times U_{23}(\psi)\times U_{13}(\phi)\times
U_{12}(\omega),$$
where $U_{ij}(\theta_{ij})$ is the mixing matrix between ith and jth
mass eigenstates with the mixing angle $\theta_{ij}$.
It has been shown that the
expression for electron neutrino survival probability, integrated over
the time of emission and of absorption, is independent of
the phase and the third Euler angle $\psi$ \cite{kujp87,ajmm88}.
They can be set to zero without loss of generality and
we have the following form for $U^v$
\begin{equation}
U^v = \left( \begin{array}{ccc}
      c_{\phi} c_{\omega} & c_{\phi} s_{\omega} & s_{\phi} \\
      -s_{\omega} & c_{\omega} & 0 \\
      -s_{\phi} c_{\omega} & -s_{\phi} s_{\omega} & c_{\phi} \\
      \end{array} \right),
\end{equation}
where $s_{\phi} = \sin \phi$ and $c_{\phi} = \cos \phi$ etc. The angles
$\omega$ and $\phi$ can take values between $0$ and $\pi/2$. Note that
one of the flavours decouples if either $\omega$ or $\phi$ is zero and we
have a two flavour scenario. As mentioned earlier the approach here is
similar to that of Joshipura and Krastev\cite{ajpik} who, however, assume that
the mixing angle between second and third generation, $\psi$, is small and
hence can be neglected.  We wish to emphasise that this is not an
assumption and infact $\psi$ can be arbitrary and the result for
survival probability of the electron neutrino is independent of this
\cite{kujp87,ajmm88}. In fact the solution of the atmospheric neutrino
deficit requires $\psi$ to be rather large.  Together, solutions of the
atmospheric neutrino deficit and the solar neutrino problem determine the
mixing matrix $U^v$ completely apart from the CP-violating phase.

The masses of the vacuum eigenstates are taken to be $\mu_1$,
$\mu_2$ and $\mu_3$.
In the mass eigenbasis, the $({\rm mass})^2$ matrix is diagonal,
\begin{eqnarray}
M_0^2  & = & \left( \begin{array}{ccc}
	             \mu_1^2 & 0 & 0 \\
                     0 & \mu_2^2 & 0 \\
	             0 & 0 & \mu_3^2  \\
		     \end{array} \right) \nonumber \\
 & = & \mu_1^2 I + \left( \begin{array}{ccc}
			 0 & 0 & 0 \\
			 0 & \delta_{21} & 0 \\
			 0 & 0 & \delta_{31} \\
			 \end{array} \right),
\end{eqnarray}
where $\delta_{21} = \mu_2^2 - \mu_1^2$ and $\delta_{31} = \mu_3^2 -
\mu_1^2$. Without loss of generality, we can take $\delta_{21}$ and
$\delta_{31}$ to be greater than zero.
Neutrino oscillation amplitudes are independent of the first
term so we drop it from further calculation. In the flavour basis
the $({\rm mass})^2$ matrix has the form
\begin{eqnarray}
%AAAA raised the subscript _v to superscript ^v
M_v^2  & = & U^v M_0^2 {U^v}^{\dagger} \nonumber \\
       & = & \delta_{31} M_{31} + \delta_{21} M_{21},
\end{eqnarray}
where
\begin{eqnarray}
M_{31} & = & \left( \begin{array}{ccc}
	s_{\phi}^2 & 0 & s_{\phi} c_{\phi} \\
	0 & 0 & 0 \\
       	s_{\phi} c_{\phi} & 0 & c_{\phi}^2 \\
         \end{array} \right) \nonumber  \\
M_{21} & = &  \left( \begin{array}{ccc}
      c_{\phi}^2 s_{\omega}^2 & c_{\phi} s_{\omega} c_{\omega} &
      -c_{\phi} s_{\phi} s_{\omega}^2 \\
      c_{\phi} s_{\omega} c_{\omega}  & c_{\omega}^2 &
      -s_{\phi} s_{\omega} c_{\omega} \\
      -c_{\phi} s_{\phi} s_{\omega}^2 & -s_{\phi} s_{\omega} c_{\omega}
       & s_{\phi}^2 s_{\omega}^2 \\ \end{array} \right).
\end{eqnarray}

As in the two flavour case, matter effects can be included by adding
$A(r)$, defined in (\ref{eq:defA}), to the $e-e$ element of $M_v^2$.
The matter
corrected $({\rm mass})^2$ matrix in the flavour basis is
\begin{equation}
M_m^2 = \delta_{31} M_{31} + \delta_{21} M_{21} + A M_A,
\label{eq:mmsq}
\end{equation}
where
\begin{equation}
M_A = \left( \begin{array}{ccc} 1 & 0 & 0 \\ 0 & 0 & 0 \\ 0 & 0 & 0 \\
       \end{array} \right).
\end{equation}
To calculate the evolution of a neutrino in matter we have to find the
matter corrected eigenstates by diagonalizing $M_m^2$. For arbitrary
values of $\delta_{31}$ and $\delta_{21}$, it is cumbersome to find the
eigenvalues and eigenvectors of $M_m^2$ algebraically. However, the
eigenvalue problem can be solved using perturbation theory, if the mass
differences have
the following hierarchy $\delta_{31} \gg \delta_{21}$. This
assumption is plausible in light of the observed atmospheric muon
neutrino deficit. Recently Kamiokande analyzed their
atmospheric neutrino data, assuming that
the deficit is caused by the oscillation of a $\nu_{\mu}$ into another
flavour. Their analysis assumes mixing between only two flavours
$(\nu_{\mu} \leftrightarrow \nu_e \ {\rm or} \ \nu_{\mu} \leftrightarrow
\nu_{\tau})$. For both cases their best fit yields a mass square
difference of the order of $10^{-2} \ {\rm eV}^2$ \cite{kamat94}.
In our analysis we take $\delta_{31}$ to be $10^{-2} \ {\rm eV}^2$.
Thus we have  $\delta_{31}$ much larger than $A_{max}$ and hence
the oscillations involving the third generation are not influenced
very much by the matter effects. In order for the matter effects to be
significant (as necessitated by the solar neutrino problem),
the other mass difference in the problem, $\delta_{21}$, should be
such that
the resonance condition is satisfied for some
values of parameters. This means
$\delta_{21} \sim A_{max}$. Thus we work in an approximation
where $\delta_{21}, A_{max} \ll \delta_{31}$.

In this approximation, to the zeroth order, both the matter term
and the term proportional to $\delta_{21}$ can be neglected in
eq.~(\ref{eq:mmsq}). Then $M_m^2 = \delta_{31} M_{31}$, whose
eigenvalues and eigenvectors are
\begin{eqnarray}
0 & ; & \left( \begin{array}{c} c_{\phi} \\ 0 \\ -s_{\phi}
	\end{array} \right), \nonumber \\
0 & ; & \left( \begin{array}{c} 0 \\ 1 \\ 0
	\end{array} \right), \nonumber \\
\delta_{31} & ; & \left( \begin{array}{c} s_{\phi} \\ 0 \\ c_{\phi}
	\end{array} \right).
\end{eqnarray}
Treating $A M_A + \delta_{21} M_{21}$ as perturbation to the
dominant term in $M_m^2$ and carrying out degenerate perturbation
theory, we get the matter dependent eigenvalues and eigenvectors,
\begin{eqnarray}
m_1^2 & ; & \left( \begin{array}{c} c_{\phi_m} c_{\omega_m} \\
	    - s_{\omega_m} \\ -s_{\phi_m} c_{\omega_m}
	    \end{array} \right), \nonumber \\
m_2^2 & ; & \left( \begin{array}{c} c_{\phi_m} s_{\omega_m} \\
	     c_{\omega_m} \\ -s_{\phi_m} s_{\omega_m}
	    \end{array} \right), \nonumber \\
m_3^2 & ; & \left( \begin{array}{c} s_{\phi_m} \\ 0 \\ c_{\phi_m}
	     \end{array} \right).
\end{eqnarray}
The above eigenvectors are the columns of the unitary matrix $U^m$
which relates the flavour eigenstates to matter dependent mass
eigenstates $\nu_i^m$ through the relation
\begin{equation}
\left[ \begin{array}{c} \nu_e \\ \nu_{\mu} \\ \nu_{\tau}
\end{array} \right] = U^m
\left[ \begin{array}{c} \nu_1^m \\ \nu_2^m \\ \nu_3^m
\end{array} \right].
\end{equation}
The matter dependent mixing angles can be expressed in terms of the
vacuum parameters and $A$ as
\begin{eqnarray}
\tan 2 \omega_m & = & \frac{\delta_{21} \sin 2 \omega}{
		  \delta_{21} \cos 2 \omega - A \cos^2 \phi},
		  \label{eq:tomegam} \\
%\tan 2 \phi_m & = & \frac{\delta_{31} \sin 2 \phi}{
%                   \delta_{31} \cos 2 \phi - A} \label{eq:tphim}.
\sin \phi_m = \sin \phi \left[1 + \frac{A}{\delta_{31}} \cos^2 \phi
\right] & ; &
\cos \phi_m = \cos \phi \left[1 - \frac{A}{\delta_{31}} \sin^2 \phi
\right]  \label{eq:phim}
\end{eqnarray}
The matter dependent eigenvalues $m_i^2$ are given by
\begin{eqnarray}
m_1^2 & = & A \cos^2 \phi \cos^2 \omega_m +
	    \delta_{21} \sin^2 \left( \omega - \omega_m \right), \nonumber \\
m_2^2 & = & A \cos^2 \phi \sin^2 \omega_m +
	    \delta_{21} \cos^2 \left( \omega - \omega_m \right), \nonumber \\
m_3^2 & = & \delta_{31} + A \sin^2 \phi \simeq \delta_{31}.
\end{eqnarray}
%Since $\delta_{31} \gg A_{max}$, $\phi_m$ will not undergo a resonance
%unless $\cos 2 \phi \ll 1$ or $\phi$ is fine-tuned to be very close to
%$\pi/4$. In this analysis, we will avoid such fine-tuned solutions and
%look, in stead, for solutions where a broad range of parameter space is
%obtainable.
$\omega_m$ can undergo a resonance if the values of
$\delta_{21}$, $\phi$ and $\omega$ are such that the resonance condition
\begin{equation}
A (r) \cos^2 \phi = \delta_{21} \cos 2 \omega \label{eq:rescon3}
\end{equation}
is satisfied for some $r$. Note that this condition is very
similar to the resonance condition in the two flavour case
(eq.~\ref{eq:rescon2}). The new feature here,
which occurs due to the mixing among the three neutrino
flavours, is the presence of the second mixing angle $\phi$ in the
resonance condition. This dependence on $\phi$ leads to a larger
region of allowed parameter space in the three flavour oscillation
scenario as will be shown in the next section. Since $\delta_{21}$,
$A(r)$ and $\cos^2 \phi$ are
all positive, a resonance can occur only if $\cos 2 \omega$
is also positive, or if $\omega < \pi/4$.

%The probability of an electron neutrino, produced in the solar core,
%being observed as an electron neutrino in one of the experiments is
In the three flavour case, the electron neutrino survival probability
is given by
\begin{equation}
\langle P_{ee} \rangle = \sum_{i,j=1}^3 \left|U^v_{ei}\right|^2
    \left|U^m_{ej}\right|^2 \left| \langle \nu^v_i \left| \right.
    \nu^m_j \rangle \right|^2. \label{eq:pee1}
\end{equation}
$\left| \langle \nu^v_i \left| \right. \nu^m_j \rangle \right|^2$
is the probability that the $j$th matter dependent eigenstate
evolves into $i$th vacuum eigenstate. As in the two flavour case,
if the adiabatic approximation holds, then
%\begin{equation}
%\frac{\delta_{21}}{E \frac{1}{A} \frac{d A}{d r}}
%\frac{\sin^2 2 \omega}{\cos 2 \omega} \ll 1
%\end{equation}
\begin{equation}
\left| \langle \nu^v_i \left| \right. \nu^m_j \rangle \right|^2
 = \delta_{ij}.
\end{equation}
We introduce the jump probabilities
\begin{equation}
x_{ij} =
\left| \langle \nu^v_i \left| \right. \nu^m_j \rangle \right|^2
\ {\rm for} \ i \neq j
\end{equation}
to take into account the non-adiabatic transitions, if the
adiabatic condition doesn't hold.

Because $\delta_{31} \gg A_{max}, \delta_{21}$, the third
eigenvalue, both in vacuum and in matter, is much larger than the
other two eigenvalues. Non-adiabatic effects are significant only if
the eigenvalues of two states come close together \cite{lanlif}.
Therefore the
jump probabilities involving the third state, $x_{13}$ and $x_{23}$
are expected to be negligibly small. Thus we have the expression for
electron neutrino survival probability to be
\begin{eqnarray}
\langle P_{ee} \rangle & = & \cos^2 \phi \cos^2 \phi_m \left(
\cos^2 \omega \cos^2 \omega_m + \sin^2 \omega \sin^2 \omega_m \right)
+ \sin^2 \phi \sin^2 \phi_m \nonumber \\
 & & - x_{12} \cos^2 \phi \cos^2 \phi_m \cos 2 \omega \cos 2 \omega_m.
 \label{eq:pee3}
\end{eqnarray}
For $x_{12}$ we use the formula,
\begin{equation}
x_{12}= \frac{\exp [ -\frac{\pi \gamma F}{2}] -\exp [ -\frac{\pi \gamma
F}{2\sin^2 \omega}]}{1 -\exp [ -\frac{\pi \gamma F}{2\sin^2 \omega}]},
\end{equation}
where $\gamma$ is defined in equation~(\ref{eq:adbcon}) and

\begin{equation}
F=1-\tan^2 \omega
\end{equation}
for an exponentially varying solar density\cite{kuopanrmp}. We use this
form  for the jump
probability  since it is valid both for large and small mixing angles.
In the extreme non-adiabatic limit $x_{12}\rightarrow \cos^2 \omega$ and
when $\gamma F >> 1$, we have the usual Landau-Zener jump probability
given by $x_{12} \rightarrow \exp[-\frac{\pi \gamma F}{2}]$ as expected.
Infact for much of the allowed parameter space, this form can be used
without any appreciable change in the results obtained.

\section{Results}
In this section we discuss the results of the numerical analysis first
for the solar neutrino problem and using that we map out the region in
the parameter space which contains the solution to  the atmospheric
neutrino problem.

\subsection{solar neutrinos}
We analyze the expression for $\langle P_{ee}\rangle$
in~(\ref{eq:pee3}) and find
the ranges of $\delta_{21}$, $\omega$ and $\phi$ allowed by the three
solar neutrino experiments. Since $\delta_{31} \gg A_{max}$, we see
%from the expression for $\tan 2\phi_m$ in~(\ref{eq:tphim}) that the
from the expression for $\phi_m$ in~(\ref{eq:phim}) that the
angle $\phi$ is almost unaffected by the matter effects. However,
$\omega_m$ can be significantly different from $\omega$ and can
undergo resonance if the resonance condition in~(\ref{eq:rescon3})
is satisfied. Since this resonance condition depends on $\phi$, in
addition to $\delta_{21}$ and $\omega$, a larger region of parameter
space satisfies the three constraints from the experiments.

To search for the regions allowed in the three parameter space
$\delta_{21}$, $\omega$ and
$\phi$, we define the suppression factors observed by the three
types of experiments
\begin{eqnarray}
y_{Ga} & = & \frac{R_{Ga;avg}}{R_{Ga;SSM}} =
             0.563 \pm 0.067, \nonumber \\
y_{Cl} & = & \frac{R_{Cl}}{R_{Cl;SSM}} =
             0.318 \pm 0.051, \nonumber \\
y_{Kam} & = & \frac{R_{Kam}}{R_{Kam;SSM}} =
             0.51 \pm 0.07, \label{eq:defyi}
\end{eqnarray}
where the first number refers to the average of the data given by two
experiments- namely GALLEX and SAGE.
The predicted SSM rates for various experiments were taken from
Bahcall-Pinsonneault SSM calculations \cite{bahpin92}. The
uncertainties in $y_i$ are the sum of the experimental uncertainty
in the numerator and the theoretical uncertainty in the
denominator, added in quadrature.

The predictions for $y_i$ for the three flavour oscillation
scenario are obtained by convoluting the SSM fluxes and the
detector cross sections with $\langle P_{ee}\rangle$
from~(\ref{eq:pee3}). The expression  we use is
\begin{equation}
y = \frac {\sum_K \int_{E_{min}}^{E_{max}} dE\Phi_K(E) \sigma(E)
<P_{ee}>(E)}{\sum_K \int_{E_{min}}^{E_{max}} dE\Phi_K(E) \sigma(E)},
\end{equation}
where the sum over K refers to the neutrino fluxes from various sources
contributing to the process. We also include the contributions
from the CNO cycle apart from the dominant contributions from the p-p
cycle.   In the case of Kamioka, only the
$^8B$ flux contributes and one must also take into account the neutral
current contribution arising from the muon neutrinos interacting with
the  detector material. The parameter ranges are then calculated by
putting vetos on $y$ at $1\sigma$ and  $1.6\sigma$ levels.
The energy dependent fluxes were taken from Ref. \cite{bahpin92} and the
cross sections were taken from Ref. \cite{bahcall}.

Figure 1 shows the
allowed values of $\omega$ and $\phi$ with $\delta_{21}$ varying between
$10^{-6} \ {\rm eV}^2$ and $10^{-4} \ {\rm eV}^2$. Note that the
allowed values of $\delta_{21}$ are also determined by the same veto
conditions. In the two generation case it is a standard practice
to plot $\delta_{21}$ against $\sin^2(2\omega)$ since that is the
combination that enters the survival probability. In the three
generation case all possible circular functions of the mixing angles
are possible. Hence we depart from the standard practice in this paper
and plot the angles themselves.   The points refer to the
allowed values after the vetos corresponding to all three experiments are
imposed. The dark squares show the values allowed by $1 \sigma$
uncertainties given in~(\ref{eq:defyi}) whereas the hollow squares show the
values allowed when the uncertainty is increased to $1.6 \sigma$.
Fig.2 shows the allowed regions in the $\phi$-$\delta_{21}$ plane, with
$\omega$ varying between $0$ and $\pi/2$ but obeying the same set of
vetos.  In Figs. 1 and 2 if we restrict ourselves to the $\phi=0$
lines (the y-axes) we get the known\cite{pgl94} two-flavour solutions
for $\omega$ and $\delta_{21}$.  The large extended regions of the
parameter space brought in through the additional degree of freedom
$\phi$ in the three-flavour scenerio are shown clearly in Figs.1 and
2.  For completeness we also plot in Fig. 3 the allowed range in
the $\omega - \delta_{21}$ plane. Here again the three-flavour
scenerio provides an enlargement of the allowed parameter space over
that of the two-flavour solution (small regions around the isolated
dark patch
in the left and around the end of the dark arm on the right).

The various regions of the
allowed parameter space may be classified as follows:
\begin{enumerate}
\item small $\delta_{21}$, small $\omega$, small $\phi$,
\item large $\delta_{21}$, large $\omega$, small $\phi$,
\item small $\delta_{21}$, small $\omega$, large $\phi$,
\item large $\delta_{21}$, small $\omega$, large $\phi$,
\item large $\delta_{21}$, large $\omega$, large $\phi$,
\end{enumerate}
where the small or large $\delta_{21}$ means either $\delta_{21} <
10^{-5} eV^2$ or $\delta_{21} > 10^{-5} eV^2$. The
first two regions corresponding to small $\phi$ in the above
classification belong to an  approximate two generation
situation  since the angle $\phi$ is small. The one corresponding to small
$\omega$ is the usual non-adiabatic solution, whereas the one
corresponding to large $\omega$  is the usual adiabatic solution. The
rest invoke the  genuine
three generation oscillation mechanism. In the two  flavour scenario, the
small   angle solution (corresponding to $\omega$ small as in case 1 above)
gives the best fit \cite{pgl94}. There the parameter space allowed at
$1 \sigma$ level is very small because the resonance condition
and the non-adiabatic jump factor fix $\delta_{21}$ and $\omega$
almost uniquely. These values of parameters indicate that the
neutrinos from the $p-p$ cycle suffer very little suppression
and those from ${}^7 Be$ suffer almost complete suppression as will be
illustrated soon in the analysis of the survival probability.

In the three flavour scenario, the resonance condition (eq.
\ref{eq:rescon3}) and the survival probability (eq.
\ref{eq:pee3}) are dependent on the second angle $\phi$ also.
The suppression of the $p-p$ neutrinos depends on the value
of $\phi$ and if this suppression is significant, then the
complete suppression for ${}^7 Be$ neutrinos can be relaxed.
This is one of the important differences between the three
flavour and  the two flavour oscillations.

Figure 4 shows the energy dependence of $\langle P_{ee}\rangle$ for
some representative
values of $\omega$, $\phi$ , and $\delta_{21}$.  The curve labelled (a)
corresponds to $\phi=2^o$.  As there is very little mixing between the
first and the third generation of neutrinos, this is infact an almost two
generation case.  In agreement with the two generation analysis, there is
almost no supression of the $p-p$ neutrinos and the $^7Be$ neutrinos are
almost completely suppressed.  The survival probability at high energies
relevant to Kamioka is almost a linear function with an average around
0.5 as one would expect.
Also here the values of $\omega$ and $\delta_{21}$
are small (they are
almost equal to the values obtained in the
two flavour case) and the non-adiabatic effects become important
beyond $2 \ {\rm MeV}$. Keeping $\omega$ small if we increase $\phi$ in
the allowed region there is a perceptible reduction in the probability in
the $p-p$ energy range and an increase in the survival probability of the
$^7Be$ neutrinos (curves (b) and (c)).  When $\delta_{21}$ is increased,
however,
there is a qualitative change in the survival probability profile.  In
this range both $\omega$ and $\phi$ are allowed to be large. Here also
there is a qualitative change when $\omega$ is small or
large. For large $\omega$ the survival probability is a smooth function
resembling the adiabatic case of the two generation analysis (curves (d)
and (f)) whereas for small $\omega$ it is almost a step function (curve
(e)) which is like the classic adiabatic case discussed by Bethe in the
two generation case
\cite{bethe86}.  One common feature of the
large $\delta_{21}$ case is that the $p-p$ neutrinos undergo
substantial suppression varying between 0.6 -0.5.  The resonance also
occurs at a much higher energy than in the small $\delta_{21}$ case.
Curve (f) has $\omega$, $\phi$ and
$\delta_{21}$  all large and in some sense it
can be called `most representative' of the three flavour
oscillation scenario because both the mixing angles in this
case are large.  In all the above cases, except (e), the average survival
probability
above 7 MeV is in the neighbourhood of 0.4 which is what is required by
the Kamioka data and there is no dramatic change from one to the other.
This is not so at low energies where the curves differ dramatically. In
this sense Kamioka experiment cannot distinguish between different
theoretical scenarios of masses and mixings.

One way of experimentally measuring the energy dependence of $\langle
P_{ee}\rangle $
is to look at the recoil electron
spectrum in those detectors that use $\nu_e -e$ scattering.  In Fig.5 we
have shown the recoil electron spectrum for the
six cases plotted in Fig. 4. Except case (f), they cannot be distinguished
beyond 10 MeV, whereas there are substantial differences at low
energies.  While this energy range is not completely accessible in
Kamioka, it is interesting to note that it may be possible to see this
difference in the experimental recoil electron spectrum  in the
SNO\cite{sno} and Borexino\cite{borex} detectors. Note that in
computing the recoil electron spectrum, we have used the spectrum of
$^8B$ neutrinos as input.  This is because  the threshold
in experiments which can measure the recoil electron spectrum (like
SNO and Kamioka) is more than a few MeVs where only this flux matters.
The only exception is Borexino where the threshold is much lower and
there are other contributions below 1.5MeV. In particular the $^7Be$
neutrino source, which is a line spectrum at 0.862 MeV, will show up
as a sharp bump
in the recoil spectrum where the height of the bump depends on the
survival probability. A complete absense of the bump
would point to the set parameters as in case (a) of Fig.4.

Finally we consider a non-standard mixing which leads  a substantial
region in the parameter space.  We consider a situation where the
electron neutrino is
coupled more strongly to the heavier mass eigenstate $\nu_2$. Obviously
this implies that the mixing angle between the first two mass
eigenstates $\omega$, is
greater than $\pi/4$.  In the standard analysis the mixing has to
be less than $\pi/4$ so that the resonance condition is satisfied as can
be seen from eq.(\ref{eq:rescon3}).  This is true in the two as well as in
the  three
generation case since the LHS of the resonance condition is positive for
arbitrary $\phi$ whereas the sign in RHS depends on the magnitude of
$\omega$.  There are no strong theoretical reasons not to consider this
situation.  Because $\omega_m$ at core is very close to $\pi/2$,
$\omega_m$ is then constrained to be $\omega \leq \omega_m <\pi/2$.
Infact for $\delta_{21} \le 10^{-7} eV^2$, $\omega_m$ is approximately
$\pi/2$. Since
$\phi$ hardly varies with density the effective survival probability may
be approximately written as,
\begin{equation}
\langle P_{ee} \rangle = \cos^4 \phi \sin^2 \omega + \sin^4\phi - x_{12}
\cos^4 \phi |\cos 2 \omega|,
\label{eq:pee4}
\end{equation}
where we have retained the jump probability $x_{12}$.  While it may
appear some what unusual to keep the jump probability when there is no
resonance, a plot of the eigenvalues clearly shows that the
difference between the first two eigenvalues  is not very different from
that of the standard case
close to vacuum and one cannot completely discard the existence of
non-adiabatic
jumps between mass eigenstates. However, most of the derivations of the
jump probability assume the existence of resonance and the profile of the
density variation close to resonance.  Since we do not have a handle on
this, we  assume that the jump probability  is simply given by
$x_{12} = \exp(-C/E)$ and treat C as a free parameter of the theory.
The survival probability is then energy dependent as would be required by
the solution to the solar neutrino puzzle.
The resulting parameter space is shown in Fig.6 for $\omega, \phi$.
The parameter C varies from 0.4 to 6.3 in the allowed region.  If we assume
any one of
the expressions for the jump probability discussed earlier, then we will
have to discard small values of C ($C<4$) since then the jump
probability becomes very large and unacceptable. However in the allowed
region, the points corresponding to small C are very few and there is no
substantial change from the plot shown in Fig.6. We also show some
typical variation of the survival probability
$\langle P_{ee}\rangle$ for some typical values
of  $\omega, \phi$ and $C$ in Fig.7. The curve (a) corresponds to
small C where non-adiabatic effects are important while the curve (b)
corresponds to large C which is an almost adiabatic case.  We wish to
stress that this is an
adhoc  solution but we have analysed this situation because there are no
strong theoretical reasons to ignore this possibility.

To conclude this section, we note that the solution to the solar neutrino
puzzle fixes  the parameter space defined by  $\omega, \phi $ and
$\delta_{21}$ .  While we have actually chosen the fourth parameter
$\delta_{31}$
to be $10^{-2} eV^2$ we  might as well have set the limit
$\delta_{31} > 10^{-3} eV^2$ without affecting our results.  One therefore
requires more inputs to fix the range of $\delta_{31}$  and the angle
$\psi$ (mixing angle between second and the third generation neutrinos)
which is arbitrary as far as the solar neutrino puzzle is
concerned. The new input is provided by the analysis of the
atmospheric neutrino problem which we consider next.

\subsection{Atmospheric Neutrinos}

In order to fix the mixing matrix completely we still need to fix the
range of $\psi$, which is the mixing angle between the second and third
generation neutrinos, as this is arbitrary in the solar neutrino
analysis. To have a consistent solution for both solar neutrino and the
atmospheric neutrino problems, we need to show that there exists a range
of $\psi$ in the allowed range of parameters occuring in the solar
neutrino problem.  To ensure this we first define the ratio
\begin{equation}
R = \frac{(\frac{\phi_{\nu_{\mu}}}{\phi{\nu_e}})_{obs}}
{(\frac{\phi_{\nu_{\mu}}}{\phi{\nu_e}})_{MC}} \label{eq.mc}
\end{equation}
which measures the ratio of the observed muon neutrino flux to the electron
neutrino flux to that expected from Monte-Carlo calculations of
neutrino production in the atmosphere. The most recent measurement of this
ratio by the Kamiokande collaboration\cite{kamat94} yields
$R=0.57^{+0.08}_{-0.07}\pm0.07$ in the multi-GeV range.  The depletion is
further confirmed by the observation of the zenith-angle dependence. The
result for sub-GeV range atmospheric neutrinos is $
R=0.60^{+0.06}_{-0.05}$ which is consistent with the multi-GeV range
data. We may therefore assume that the suppression is approximately
energy independent from sub-GeV to multi-GeV range of energies.
Assuming that this depletion is due to the vacuum oscillations amongst
the neutrino flavours, this ratio may be written as,
\begin{equation}
R = \frac {P_{\mu\mu} + r P_{e\mu}}{P_{ee}+\frac{1}{r}P_{\mu e}},
\label{ratm}
\end{equation}
where $r=\phi_{\nu_e}/\phi_{\nu_\mu}=0.45$ is the ratio of the flux of
electron  neutrinos to that of muon neutrinos at the production point.
Note that $r$ is simply the inverse of the flux ratio expected on the
basis of the Monte-Carlo calculations (see eq.(\ref{eq.mc})).
We now assume that the survival probabilities ($P_{ee},P_{\mu\mu}$) and the
oscillation probability ($P_{\mu e}$) are given by the full three
generation mixing matrix defined by the angles $\phi, \omega , \psi$ and
the two mass squared differences $\delta_{21}, \delta_{31}$.
The vacuum oscillation probability between two flavours is then given by,
\begin{eqnarray}
&&P_{ij} = U_{i1}^2~U_{j1}^2+ U_{i2}^2~U_{j2}^2+ U_{i3}^2~U_{j3}^2
+2U_{i1}U_{i2}U_{j1}U_{j2}\cos(2.53\frac{d \delta_{21}}{E})\\
\nonumber
&&+2U_{i1}U_{i3}U_{j1}U_{j3}\cos(2.53\frac{d \delta_{31}}{E})
+2U_{i3}U_{i2}U_{j3}U_{j2}\cos(2.53 \frac{d \delta_{32}}{E}),
\end{eqnarray}
where i and j are the flavour indices,  E is the energy given in units
of MeV, $\delta_{ij}$ is the mass differences in $eV^2$ and $d$ is the
distance of traversal given  in  meters.
These probabilities explicitly depend on the distance $d$ travelled by the
neutrinos from the point of production to the point of detection and
is approximately about 13,000 kms for the upward
moving neutrinos. This
distance is much less than the oscillation length between the first two
generations( since $\delta_{21}$ is small). Therefore the cosine factor
involving
$\delta_{21}$ can be safely set equal to unity.  As mentioned before,
Kamioka  has also observed
that the level of suppression for the atmospheric muon neutrinos is
approximately the same both for sub-GeV and the multi-GeV neutrinos. This
can be ensured if the energy dependent factors involving $\delta_{31}$
and $\delta_{32}$ are such that the cosine functions can be replaced by
the corresponding averages. This is possible if and only if many
oscillation lengths are contained in the distance travelled by neutrinos
to the detector. This then sets the limits on the mass squared difference
$\delta_{31} > 10^{-3} eV^2$.  The large  $\delta_{31}$ regions
($\delta_{31}>10^{-1} eV^2$) are excluded at 90 percent C.L by the analysis
of the multi-GeV neutrino data\cite{kamat94}.   While we have used the
central value $10^{-2} eV^2$ in our solar neutrino analysis, the results
for both solar and atmospheric neutrinos
remain unchanged if the value is further increased and marginal
changes occur for values close to $10^{-3} eV^2$ because of the
approximations we made in the solar neutrino analysis.

Therefore the only
range to be fixed is for the mixing angle $\psi$.  This we do by
requiring the theoretical value of R  calculated from eq.(\ref{ratm})
is within $1\sigma$ and
$1.6\sigma$ of the experimental value.  The resulting range for $\psi$ is
shown in Fig.8, whereas usual the full squares show the $1\sigma$
veto and the open squares show $1.6\sigma$ veto.

A few comments are in order here:  As in the two generation analysis of
the atmospheric neutrino problem, we find that the preferred values of
$\psi$ is large and around $\pi/4$.  This can be checked easily by looking
at the conversion probability $P_{\mu e}$ in the allowed range of
parameters for the atmospheric neutrino problem.  It turns out that this
conversion probability is always less than twenty percent.  Thus the
solution to the atmospheric neutrino problem is mainly driven by the
$\nu_\mu-\nu_\tau$ oscillations whereas the solution to the solar
neutrino problem is mainly driven by $\nu_e- \nu_\mu$ oscillations at
least for small values of $\phi$. However there are large domains of the
parameter space where one requires the full three generation analysis
presented here, to have a consistent solution to both the problems.

\section{Summary and Discussion}
We have examined in detail the possible solutions to the solar neutrino
and atmospheric neutrino puzzles in the realistic three generation
framework.  There are in general three mixing angles, one
phase from the mixing matrix and two mass squared differences which
define the full parameter space. In the case of solar neutrinos the
survival probability for the
electron neutrino, even after taking into account the matter effects, is
independent of the phase and one of the mixing angles. We also fix one of
the mass squared differences by appealing to the atmospheric neutrino
problem. Thus our parameter space in the solar neutrino analysis consists
of  two angles and one mass
squared difference. In our case these are chosen to be $\omega$ which
gives the
mixing between first and second generations , $\phi$ which is the mixing
between first and third generations and $\delta_{21}$ which is the mass
squared difference between the first two generations. The mass
difference $\delta_{31}$ is fixed to be around $10^{-2} eV^2$ to explain
the atmospheric neutrino problem.  We have mapped out the
parameter space($\phi, \omega, \delta_{21}$) by invoking the vetos
arising from the data given by  the
three solar neutrino experiments. Next we have used these allowed ranges of
parameters
from the solar neutrino analysis as input in the atmospheric neutrino
analysis to fix the angle $\psi$ and find that there exists a substantial
range in this parameter which allows a solution to the atmospheric
neutrino puzzle.  The numerical calculations necessarily depend on the
bin size for the parameters. We have ensured that the bin size we have
chosen is such that a further reduction will not change the overall
profile of the allowed region.  However it is conceivable that the
rough edges that one still sees in parts of the allowed region will be
smoothed out by a further reduction of the bin size.

In conclusion, we have shown that there exists a consistent solution to
the solar and atmospheric neutrino deficit puzzles within the framework
of standard MSW mechanism based on the set of all available
measurements of the solar neutrino fluxes. The full analysis
involves five parameters which we have mapped out by accommodating the
solar and atmospheric neutrino fluxes seen by the present set of
experiments.  While the allowed region in the parameter
space is still large, these can be constrained further by measuring
the distributions of recoil
electron  energies in solar neutrino detectors that use $\nu - e$
scattering.  Although the threshold energy at the Kamioka detector is
rather too high for this purpose, the  SNO and Borexino detectors may
be effective in narrowing the parameter space. Finally we would like
to remark that the  analysis of solar and atmospheric neutrino
problems presented here is exploratory in nature.  This is so since with
time the errors are bound to change which inturn will affect the vetos
imposed by us at $1\sigma$ and $1.6\sigma$ levels. Nevertheless we
believe there is already  sufficient indication that a robust
solution of both problems is possible within the framework provided by
the mechanism of neutrino oscillations with three generations.

Acknowledgements: This work was started during the Third Workshop on
High Energy Particle Physics (WHEPP 3) in Madras. We gratefully
acknowledge the  collaboration in the initial stages with
Prof.A.S.Joshipura.  We also thank Prof. K.V.L.Sarma for keeping us
informed of the recent developments in  neutrino oscillations.

\newpage

\begin{figure}
\caption{
Allowed regions in $\phi - \omega$ plane $( {\rm with} \
10^{-6}  \ {\rm eV}^2 \leq \delta_{21} \leq 10^{-4} \ {\rm
eV}^2)$ at $1 \sigma$ (dark squares) and at $1.6 \sigma$
(hollow squares).
}\label{fig1a}
\end{figure}

\begin{figure}
\caption{
Allowed regions in $\phi - \log (\delta_{21}/{\rm eV}^2)$ plane
$( \rm with \ 0 \leq \omega \leq \pi/2)$ at $1 \sigma$ (dark
squares) and at $1.6 \sigma$ (hollow squares).
}\label{fig1b}
\end{figure}

\begin{figure}
\caption{
Allowed regions in $\omega - \log (\delta_{21}/{\rm eV}^2)$ plane
$( \rm with \ 0 \leq \phi \leq \pi/2)$ at $1 \sigma$ (dark
squares) and at $1.6 \sigma$ (hollow squares).
}\label{fig1c}
\end{figure}

\begin{figure}
\caption{
Survival probability
$\langle P_{ee}\rangle$ vs $E_{\nu}$ for typical values of $\phi$,
$\omega$ and
$\delta_{21}$ in the allowed region. The parameters chosen are:
(a) $\delta_{21} = 4.0\times 10^{-6}, \omega =2.5^o, \phi = 2.0^o$;
(b) $\delta_{21} = 5.0\times 10^{-6}, \omega =2.0^o, \phi = 16.5^o$;
(c) $\delta_{21} = 7.0\times 10^{-6}, \omega =1.75^o, \phi =37.5^o$;
(d) $\delta_{21} = 2.5\times 10^{-5}, \omega =35.0^o, \phi = 3.0^o$;
(e) $\delta_{21} = 7.0\times 10^{-5}, \omega =2.0^o, \phi = 30.0^o$;
(f) $\delta_{21} = 1.0\times 10^{-4}, \omega =24.5^o, \phi =24.0^o$;
$\delta_{21}$ is given in terms of ${\rm eV}^2$.
}\label{fig2}
\end{figure}

\begin{figure}
\caption{
Recoil electron spectrum for different representative points of the
allowed parameter region.  The parameters for the differenct curves
labelled (a)-(f) are the same as in Fig.4. The inset shows a
comparison of all zix cases with the SSM spectrum(dashed line).
}\label{fig3}
\end{figure}

%\begin{figure}
%\caption{
%Eigenvalue spectrum for the standard case, where all the angles are
%less than $\pi/4$ (shown by the solid line) and for the non-standard case
%(see text) shown by the dashed line. Note that there
%is resonance between the first two generations in the first case while
%the condition is not satisfied in the second case. Only the first two
%eigenvalues are shown. The third is almost decoupled and is not shown in
%this figure.
%}\label{fig4}
%\end{figure}

\begin{figure}
\caption{
Allowed regions in $\phi - \omega$ plane $(
{\rm with}  0.4\le C \le 6.4$ (dark squares) and at $1.6 \sigma$
(hollow squares) for the non-standard solutions.
}\label{fig5}
\end{figure}

\begin{figure}
\caption{
Typical survival probability profile in the non-standard case.  The
curve labelled (a) corresponds to $C=0.4,\omega=55^o, \phi=2^o$ and
the curve labelled (b) corresponds to $C=6,\omega=89^o, \phi=38^o$.
}\label{fig6}
\end{figure}

\begin{figure}
\caption{
The allowed range of values for the mixing angle $\psi$  in the $\psi -
\phi $ plane when the $\phi$
and $\omega$ are restricted to the range allowed by the solar neutrino
problem(see Fig.1).
}\label{fig7}
\end{figure}

\end{document}